\documentclass[journal]{IEEEtran}
\IEEEoverridecommandlockouts
\usepackage{amsmath}
\usepackage{amsfonts}
\usepackage{amsthm} 
\usepackage{amssymb} 
\usepackage{breqn}
\usepackage{setspace}
\usepackage[usenames,dvipsnames]{xcolor}
\usepackage{graphicx}
\usepackage{endnotes}
       \usepackage{sidecap}
       \sidecaptionvpos{figure}{c}
\usepackage[T1]{fontenc}
\usepackage{supertabular}
\usepackage{longtable}
\usepackage{bbm}

\usepackage{mdwmath}
\usepackage{mdwtab}
\title{\color{Brown} Tail Risk Constraints and Maximum Entropy}
\author{Donald Geman, H\'elyette Geman, and Nassim Nicholas Taleb}

\author{
    \IEEEauthorblockN{Donald Geman\IEEEauthorrefmark{1}, H\'elyette Geman\IEEEauthorrefmark{1}\IEEEauthorrefmark{2}, and Nassim Nicholas Taleb \IEEEauthorrefmark{3}}\\
    
    \IEEEauthorblockA{\IEEEauthorrefmark{1}Dept of Applied Mathematics,
Johns Hopkins University    \\}
    \IEEEauthorblockA{\IEEEauthorrefmark{2}Dept of Mathematics, 
Birkbeck, University of London     \\} \IEEEauthorblockA{\IEEEauthorrefmark{2}School of Engineering, New York University} 
}

\date{December 2014}

\begin{document}
\maketitle
\setstretch{1.1}
\begin{abstract} 
In the world of modern financial theory, portfolio construction has
traditionally operated under at least one of two central assumptions:
the constraints are derived from a utility function and/or the
multivariate probability distribution of the underlying asset returns
is fully known.  In practice, both the performance criteria and the
informational structure are markedly different: risk-taking agents are mandated
to build portfolios by primarily constraining the tails of the
portfolio return to satisfy VaR, stress testing, or expected shortfall
(CVaR) conditions, and are largely ignorant about the
remaining properties of the probability distributions.  As an
alternative, we derive the shape of portfolio distributions which have
maximum entropy subject to real-world left-tail constraints and other
expectations.  Two consequences are (i) the left-tail constraints are
sufficiently powerful to overide other considerations in the
conventional theory, rendering individual portfolio components of
limited relevance; and (ii) the "barbell" payoff (maximal
certainty/low risk on one side, maximum uncertainty on the other)
emerges naturally from this construction.
\end{abstract}

\maketitle 

\thispagestyle{empty}

\section{Left Tail Risk as the Central Portfolio Constraint}

Customarily, when working in an institutional framework, operators and
risk takers principally use regulatorily mandated tail-loss limits to
set risk levels in their portfolios (obligatorily for banks since
Basel II).  They rely on stress tests, stop-losses, value at risk
(VaR), expected shortfall (CVaR), and similar loss curtailment
methods, rather than utility.\footnote{In particular the margining of
financial transactions (which allows a certain leverage and targets
position size) is calibrated by clearing firms and exchanges on tail
losses, seen both probabilistically and through stress testing.} The
information embedded in the choice of the constraint is, to say the
least, a meaningful statistic about the appetite for risk and the
shape of the desired distribution.

Operators are less concerned with portfolio variations than with the drawdown they may face over a time window. Further, they are in ignorance of
the joint probability distribution of the components in their portfolio (except for a vague notion of association and hedges), but can control losses organically with
allocation methods based on maximum risk.\footnote{The idea of
substituting variance for risk
can appear very strange to
practitioners of risk-taking. The aim by Modern Portfolio Theory at
lowering variance is inconsistent with the preferences of a rational
investor, regardless of his risk aversion, since it also minimizes the
variability in the profit domain --except in the very narrow situation
of certainty about the future mean return,
and in the far-fetched case where the investor can only invest in
variables having a symmetric probability distribution, and/or only
have a symmetric payoff. Stop losses and tail risk controls violate
such symmetry.}

The conventional notions of utility and variance may be used, but not
directly as information about them is embedded in the tail loss constaint. 

Since the stop loss, the VaR (and expected shortfall) approaches and
other risk-control methods concern only one segment of the
distribution, the negative side of the loss domain, we can get a dual
approach akin to a portfolio separation, or "barbell-style"
construction, as the investor can have opposite stances on different
parts of the return distribution. Our definition of barbell here is
the mixing of two extreme properties in a portfolio such as a linear
combination of maximal conservatism for a fraction $w$ of the portfolio,
with $w \in (0,1)$, on one hand and maximal (or high) risk on the
$(1-w)$ remaining fraction.

Historically, finance theory has had a preference for parametric, less
robust, methods.  The idea that a decision-maker has clear and error-free knowledge about the distribution of future payoffs has survived in spite of its lack of
practical and theoretical validity --for instance, correlations are too unstable to yield precise measurements.\footnote{Correlations are unstable in an
unstable way, as joint returns for assets are not elliptical, see Bouchaud and Chicheportiche (2012)  
\cite{chicheportiche2012joint}.} It is an approach that is based on
distributional and parametric certainties, one that may be useful for
research but does not accommodate responsible risk taking.

There are roughly two traditions: one based on highly parametric
decision-making by the economics establishment (largely represented by
Markowitz \cite{markowitz1952portfolio}) and the other based on somewhat
sparse assumptions and known as the Kelly criterion.\footnote{Kelly,
1956 \cite{kelly1956new}, see Bell and Cover, 1980
\cite{bell1980competitive}. In contrast to the minimum-variance
approach, Kelly's method, developed around the same period as
Markowitz, requires no joint distribution or utility function. In
practice one needs the ratio of expected profit to worst-case return
dynamically adjusted to avoid ruin. Obviously, model error is of smaller consequence under the Kelly criterion: Thorp
(1969)\cite{thorp1969optimal}, Haigh (2000) \cite{haigh2000kelly}, Mac
Lean, Ziemba and Blazenko \cite{maclean1992growth}.  For a discussion
of the differences between the two approaches, see Samuelson's objection
to the Kelly criterion and logarithmic sizing in Thorp 2010
\cite{thorp2010understanding}.} Kelly's method is also related to left-
tail control due to proportional investment, which automatically
reduces the portfolio in the event of losses; but the original method
requires a hard, nonparametric worst-case scenario, that is, securities
that have a lower bound in their variations, akin to a gamble in a
casino, which is something that, in finance, can only be accomplished through
binary options. The Kelly criterion, in addition, requires some
precise knowledge of future returns such as the mean. Our approach
goes beyond the latter method in  accommodating more uncertainty about the
returns, whereby an operator can only control his
left-tail via derivatives and other forms of insurance or dynamic
portfolio construction based on stop-losses.

In a nutshell, we hardwire the curtailments on loss but otherwise assume
maximal uncertainty about the returns.  More precisely,
we equate the return distribution with the {\it maximum
entropy extension} of constraints expressed as statistical
expectations on the left-tail behavior as well as on the expectation
of the return or log-return in the non-danger zone.  
Here, the ``left-tail behavior'' refers
to the hard, explicit, institutional constraints discussed above. We describe the shape and
investigate other properties of the resulting so-called {\it maxent}
distribution. In addition to a mathematical result revealing the link
between acceptable tail loss (VaR) and the expected return in the
Gaussian mean-variance framework, our contribution is then twofold: 1)
an investigation of the shape of the distribution of returns from
portfolio construction under more natural constraints than those
imposed in the mean-variance method, and 2) the use of stochastic
entropy to represent residual uncertainty.

VaR and CVaR methods are not error free --parametric VaR is known to
be ineffective as a risk control method on its own. However, these methods can be made
robust using constructions that, upon paying an insurance price, no
longer depend on parametric assumptions. This can be done using
derivative contracts or by organic construction (clearly if someone
has 80\% of his portfolio in num\'eraire securities, the risk of
losing more than 20\% is zero independent from all possible models of
returns, as the fluctuations in the num\'eraire are not considered
risky). We use "pure robustness" or both VaR and zero shortfall via
the "hard stop" or insurance, which is the special case in our paper
of what we called earlier a "barbell" construction.\footnote{It is
worth mentioning that it is an old idea in economics that an investor
can build a portfolio based on two distinct risk categories, see Hicks
(1939). Modern Portfolio Theory proposes the mutual fund theorem or
"separation" theorem, namely that all investors can obtain their
desired portfolio by mixing two mutual funds, one being the riskfree
asset and one representing the optimal mean-variance portfolio that is
tangent to their constraints; see Tobin
(1958) \cite{tobin1958liquidity}, Markowitz (1959) \cite{markowitz1959}, and the variations
in Merton (1972) \cite{merton1972analytic}, Ross 
(1978) \cite{ross1978mutual}. In our case a riskless asset is the part
of the tail where risk is set to exactly zero. Note that the risky
part of the portfolio needs to be minimum variance in traditional
financial economics; for our method the exact opposite representation
is taken for the risky one.}

\subsection*{The Barbell as seen by E.T. Jaynes}

Our approach to constrain only what can be constrained (in a robust
manner) and to maximize entropy elsewhere echoes a remarkable insight by
E.T. Jaynes in "How should we use entropy in
economics?" \cite{jaynesy1991should}:
\begin{quotation}
It may be that a macroeconomic system does not move in response to (or
at least not solely in response to) the forces that are supposed to
exist in current theories; it may simply move in the direction of
increasing entropy as constrained by the conservation laws imposed by
Nature and Government.
\end{quotation}

\section{Revisiting the Mean Variance Setting}
Let $\vec{X}=(X_1,...,X_m)$ denote $m$ asset returns over a given
single period with joint density $g(\vec{x})$,
mean returns $\vec{\mu}=(\mu_1,...,\mu_m)$ and $m \times m$ covariance
matrix $\Sigma$: $\Sigma_{ij}= \mathbb{E}(X_iX_j)-\mu_i \mu_j, 1 \leq i,j \leq
m$.  Assume that $\vec{\mu}$ and $\Sigma$ can be reliably
estimated from data.

The return on the portolio with weights
$\vec{w}=(w_1,...,w_m)$ is then
\[ X = \sum_{i=1}^m w_iX_i,\]
which has mean and variance
\[  \mathbb{E}(X)=\vec{w}\vec{\mu}^T,\,\,\,\,\, V(X)=\vec{w}\Sigma \vec{w}^T.\]
In standard portfolio theory one minimizes $V(X)$ over all $\vec{w}$ subject
to $\mathbb{E}(X)=\mu$ for a fixed desired average return $\mu$.  Equivalently,
one maximizes the expected return $\mathbb{E}(X)$ subject to a fixed variance $V(X)$.
In this framework variance is taken as a substitute for risk.

To draw connections with our entropy-centered approach, we consider two standard cases:
\begin{enumerate}
\item {\bf Normal World:} The joint distribution $g(\vec{x})$ of asset returns is
multivariate Gaussian $N(\vec{\mu},\Sigma)$.  Assuming normality is
equivalent to assuming $g(\vec{x})$ has maximum (Shannon) entropy
among all multivariate distributions with the given first- and
second-order statistics $\vec{\mu}$ and $\Sigma$.  Moreover, for a
fixed mean $\mathbb{E}(X)$, minimizing the variance $V(X)$ is equivalent to
minimizing the entropy (uncertainty) of $X$.  (This is true since
joint normality implies that $X$ is univariate normal for any choice
of weights and the entropy of a $N(\mu,\sigma^2)$ variable is
$H=\frac{1}{2}(1+\log(2\pi\sigma^2))$.) This is natural in a world
with complete information.\footnote{ The idea of entropy as mean
uncertainty is in Philippatos and Wilson (1972) 
\cite{philippatos1972entropy}; see Zho et al. (2013)  
\cite{zhou2013applications} for a review of entropy in financial
economics and Georgescu-Roegen (1971)
\cite{georgescu1971entropy} for economics in general. }
\item {\bf Unknown Multivariate Distribution:}  
Since we assume we can estimate the second-order structure, we 
can still carry out the
Markowitz program, i.e., choose the portfolio weights to find an optimal
mean-variance performance, which determines $\mathbb{E}(X)=\mu$ and $V(X)=\sigma^2$.
However, we do not know the distribution of the
return $X$.  Observe that {\it assuming} $X$ is normal $N(\mu,\sigma^2)$
is equivalent to assuming the entropy of $X$ is {\it maximized} since,
again, the normal maximizes entropy at a given mean and variance, see
\cite{philippatos1972entropy}.
\end{enumerate}

Our strategy is to generalize the second scenario by replacing the
variance $\sigma^2$ by two left-tail value-at-risk constraints and to
model the portfolio return as the maximum entropy extension of these
constraints together with a constraint on the overall performance or
on the growth of the portfolio in the non-danger zone.

\subsection*{Analyzing the Constraints}

Let $X$ have probability density $f(x)$.  In everything that follows,
let $K<0$ be a normalizing constant chosen to be consistent with the
risk-taker's wealth.  For any $\epsilon > 0$ and $\nu_- < K$, the {\it
value-at-risk constraints} are:
\begin{enumerate}
\item {\bf Tail probability:}
\[ \mathbb{P}(X \leq K) = \int_{-\infty}^K f(x)\,\mathrm{d}x =    \epsilon.\]
\item {\bf Expected shortfall (CVaR):}
\[ \mathbb{E}(X|X \leq K) = \nu_-.\]
\end{enumerate}
Assuming 1) holds, constraint 2) is equivalent to
\[ \mathbb{E}(XI_{(X \leq K)})=\int_{-\infty}^K xf(x)\,\mathrm{d}x = \epsilon \nu_-. \]
Given the value-at-risk parameters $\theta=(K,\epsilon,\nu_-)$,
let $\Omega_{var}(\theta)$ denote the set of probability densities $f$ satisfying
the two constraints.  Notice that $\Omega_{var}({\theta})$ is convex: $f_1,f_2 \in
\Omega_{var}(\theta)$ implies $\alpha f_1 + (1-\alpha)f_2 \in 
\Omega_{var}(\theta).$
Later we will add another constraint involving the
overall mean.
\begin{figure}
\includegraphics[width=\linewidth]{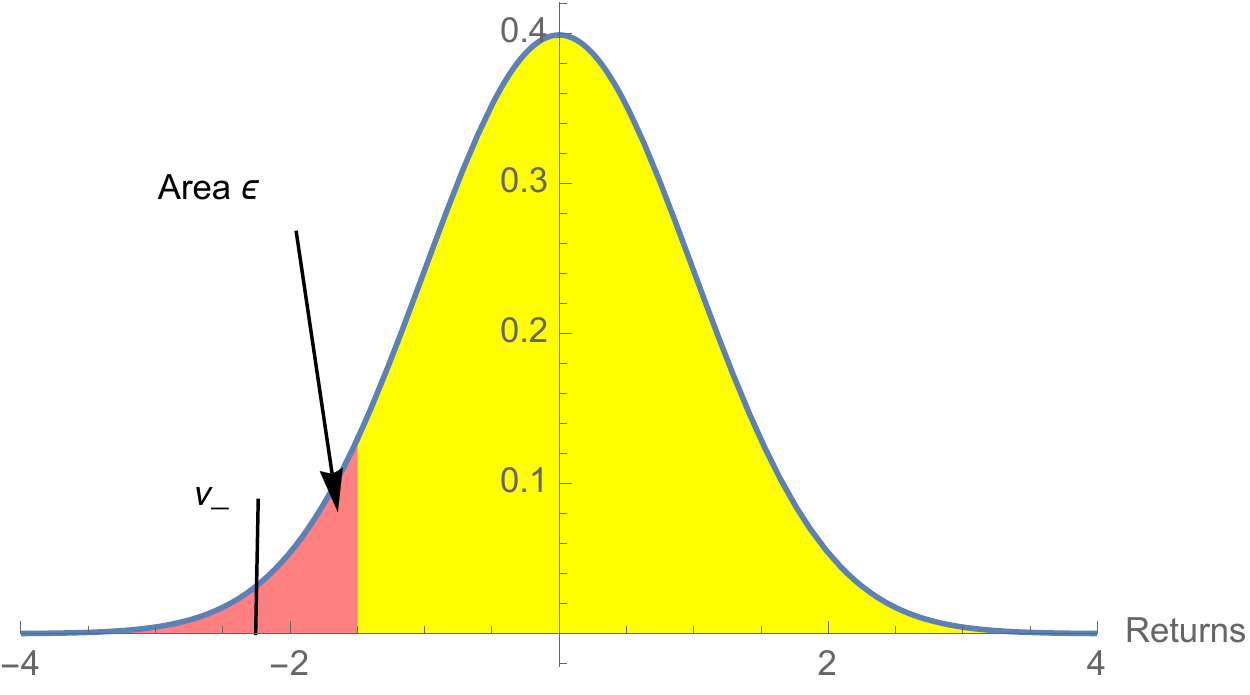}
\caption{By setting K (the value at risk), the probability $\epsilon$ of
exceeding it, and the shortfall when doing so, there is no wiggle room
left under a Gaussian distribution: $\sigma$ and $\mu$ are determined,
which makes construction according to portfolio theory
less relevant.}\label{simplefigure}
\end{figure}

\section{Revisiting the Gaussian Case}\label{restrictgaussian}

Suppose we assume $X$ is Gaussian with mean $\mu$ and variance
$\sigma^2$.  In principle it should be possible to satisfy
the VaR constraints since we have two free parameters.
Indeed, as shown below, the left-tail constraints determine the
mean and variance; see Figure \ref{simplefigure}.
However, satisfying the VaR constraints imposes interesting
restrictions on $\mu$ and $\sigma$ and leads to a natural
inequality of a "no free lunch" style.

Let $\eta(\epsilon)$ be the $\epsilon$-quantile of the
standard normal distribution, i.e., $\eta(\epsilon)=
\Phi^{-1}(\epsilon)$, where $\Phi$ is the c.d.f. of the
standard normal density $\phi(x)$.
In addition, set
\[  B(\epsilon)=\frac{1}{\epsilon \eta(\epsilon)}\phi(\eta(\epsilon))
= \frac{1}{\sqrt{2\pi}\epsilon \eta(\epsilon)} 
\exp\{-\frac{\eta(\epsilon)^2}{2}\}.\]

\bigskip

\noindent
{\bf Proposition 1:} {\it If $X \sim N(\mu,\sigma^2)$ and satisfies
the two VaR constraints, then the mean and variance are given by:}
\[  \mu = \frac{\nu_- + KB(\epsilon)}{1+B(\epsilon)},\,\,\,
\sigma = \frac{K-\nu_-}{\eta(\epsilon)(1+B(\epsilon))}.\]
{\it Moreover, $B(\epsilon) < -1$ and $\lim_{\epsilon \downarrow 0}B(\epsilon)
=-1$.}

\bigskip

The proof is in the Appendix.  The VaR constraints
lead directly to two linear equations in $\mu$ and $\sigma$:
\[  \mu + \eta(\epsilon)\sigma = K,\,\,\,\, \mu - 
\eta(\epsilon)B(\epsilon)\sigma = \nu_-.\]

Consider the conditions under which the
VaR constraints allow a \textit{positive} mean return $\mu=\mathbb{E}(X) >0$.
First, from the above linear equation
in $\mu$  and $\sigma$ in terms of $\eta(\epsilon)$ and $K$, we see
that $\sigma$ increases as $\epsilon$ increases for any fixed
mean $\mu$, and that $\mu>0$ if and only if
$\sigma > \frac{K}{\eta(\epsilon)}$, i.e., we must accept a lower
bound on the variance which increases with $\epsilon$, which is a reasonable property.  Second, from the expression for $\mu$ in Proposition 1, we have
\[  \mu > 0 \iff |\nu_-| > KB(\epsilon).\]
Consequently, the only way to have a positive expected return is to
accommodate a sufficiently large risk expressed by the various
tradeoffs among the risk parameters $\theta$ satisfying the inequality
above. \footnote{ This type of restriction also applies more generally
to symmetric distributions since the left tail constraints impose a
structure on the mean and scale. For instance, in the case of a
Student T distribution with scale $s$, location $m$, and tail exponent
$\alpha$, the same linear relation between $s$ and $m$ applies: $s =
(K-m) \kappa(\alpha)$, where $\kappa(\alpha)= -\frac{i \sqrt{I_{2
\epsilon }^{-1}\left(\frac{\alpha
}{2},\frac{1}{2}\right)}}{\sqrt{\alpha } \sqrt{I_{2 \epsilon
}^{-1}\left(\frac{\alpha }{2},\frac{1}{2}\right)-1}}$, where $I^{-1}$
is the inverse of the regularized incomplete beta function $I$, and $s$ the
solution of $\epsilon =\frac{1}{2} I_{\frac{\alpha s ^2}{(k-m
)^2+\alpha s^2}}\left(\frac{\alpha }{2},\frac{1}{2}\right)$. 
}

\subsection*{A Mixture of Two Normals}

In many applied sciences, a mixture of two normals provides a useful
and natural extension of the Gaussian itself; in finance, the Mixture
Distribution Hypothesis (denoted as MDH in the literature) refers to a
mixture of two normals and has been very widely investigated (see for
instance Richardson and Smith (1995) 
\cite{richardson1994direct}). H. Geman and T.An\'e (1996) 
\cite{ane2000order} exhibit how an infinite mixture of normal distributions for stock returns arises from the introduction of a "stochastic clock" accounting for
the uneven arrival rate of information flow in the financial markets.
In addition, option traders have long used mixtures to account for fat
tails, and to examine the sensitivity of a portfolio to an increase in
kurtosis ("DvegaDvol"); see Taleb (1997) 
\cite{taleb1997dynamic}. Finally, Brigo and Mercurio (2002) 
\cite{brigo2002lognormal} use a mixture of two normals to calibrate the
skew in equity options.

Consider the mixture
\[  f(x)=\lambda N(\mu_1,\sigma_1^2) + (1-\lambda) N(\mu_2,\sigma_2^2).\]
An intuitively simple and appealing case is to fix the overall mean
$\mu$, and take $\lambda=\epsilon$ and $\mu_1=\nu_-$, in which case
$\mu_2$ is constrained to be $\frac{\mu - \epsilon \nu_-}{1-\epsilon}$.  
It then follows
that the left-tail constraints are approximately satisfied for $\sigma_1,\sigma_2$
sufficiently small.  Indeed, when $\sigma_1=\sigma_2 \approx 0$,
the density is effectively composed of two spikes
(small variance normals) with the left one
centered at $\nu_-$ and the right one centered at
at $\frac{\mu - \epsilon \nu_-}{1-\epsilon}$.
The extreme case is a Dirac function
on the left, as we see next.

\subsubsection*{Dynamic Stop Loss, A Brief Comment}
\begin{figure}
\includegraphics[width=\linewidth]{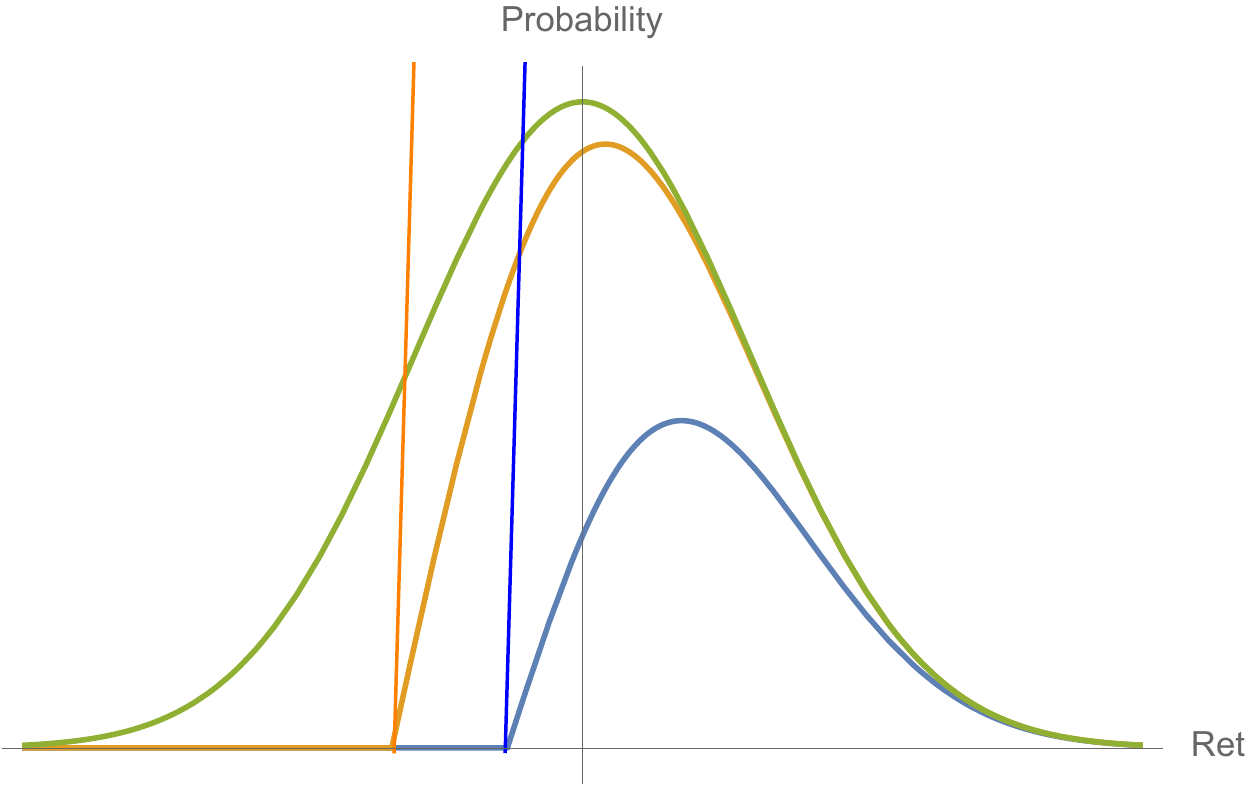}
\caption{Dynamic stop loss acts as an absorbing barrier, with a Dirac function at the executed stop.}\label{dynamicstoploss}
\end{figure}

One can set a level $K$ below which there is no mass, with
results that depend on accuracy of the execution of such a stop. The
distribution to the right of the stop-loss no longer looks like the
standard Gaussian, as it builds positive skewness in accordance to the
distance of the stop from the mean. We limit any further discussion 
to the illustrations in Figure
\ref{dynamicstoploss}.

\section{Maximum Entropy}

From the comments and analysis above, it is clear that, in practice,
the density $f$ of the return $X$ is unknown; in particular, no theory
provides it.  Assume we can adjust the portfolio parameters to satisfy
the VaR constraints, and perhaps another constraint on the expected
value of some function of $X$ (e.g., the overall mean).  We then wish
to compute probabilities and expectations of interest, for example
$\mathbb{P}(X>0)$ or the probability of losing more than $2K$, or the expected
return given $X>0$. One strategy is to make such estimates and
predictions under the most unpredictable circumstances consistent with
the constraints.  That is, use the {\it maximum entropy
extension} (MEE) of the constraints as a model for $f(x)$.

The ``differential entropy'' of $f$ is $h(f) = -\int f(x) \ln f(x)
\,\mathrm{d}x$.  (In general, the integral may not exist.)  Entropy is concave on
the space of densities for which it is defined as:
\[ h(\alpha f_1+(1-\alpha)f_2) \geq \alpha h(f_1) + (1-\alpha) h(f_2).\]
We want to maximize entropy subject to the VaR constraints together
with any others we might impose.  Indeed, the VaR constraints alone do
not admit an MEE since they do not restrict the density $f(x)$ for
$x>K$.  The entropy can be made arbitrarily large by allowing $f$ to
be identically $C=\frac{1-\epsilon}{N-K}$ over $K<x<N$ and letting $N
\rightarrow \infty$.  Suppose, however, that we have adjoined one or
more constraints on the behavior of $f$ which are compatible with the
VaR constraints in the sense that the set of densities $\Omega$
satisfying all the constraints is non-empty. Here $\Omega$ would
depend on the VaR parameters $\theta=(K,\epsilon,\nu_-)$ together with
those parameters associated with the additional constraints.  The MEE
is then defined as
\[  f_{MEE} = \arg \max_{f \in \Omega} h(f).\]
It is known that $f_{MEE}$ is unique and (away from the
boundary of feasibility) is an exponential distribution in the
constraint functions.  

\subsection{Case A: Constraining the Global Mean}

The simplest case is to add a constraint on the mean return, i.e.,
fix $\mathbb{E}(X)=\mu$. Since $\mathbb{E}(X)=\mathbb{P}(X \leq K)\mathbb{E}(X|X\leq K)+\mathbb{P}(X>K)\mathbb{E}(X|X>K)$, adding the mean
constraint is equivalent to adding the constraint 
\[  \mathbb{E}(X|X>K)=\nu_+ \]
where $\nu_+$ satisfies $\epsilon \nu_- + (1-\epsilon) \nu_+ = \mu$.

Define

\begin{equation*}
f_-(x) =
\begin{cases}
\frac{1}{(K-\nu_-)} \exp 
\left[ -\frac{K-x}{K-\nu_-} \right] & \text{if } x < K,\\
0 & \text{if } x \geq K.
\end{cases}
\end{equation*}

and

\begin{equation*}
f_+(x) =
\begin{cases}
\frac{1}{(\nu_+-K)} \exp 
\left[ -\frac{x-K}{\nu_+-K} \right] & \text{if } x > K,\\
0 & \text{if } x \leq K.
\end{cases}
\end{equation*}

It is easy to check that both $f_-$ and $f_+$ integrate to one.  Then
\[  f_{MEE}(x) = \epsilon f_{-}(x) + (1-\epsilon) f_{+}(x)\]
is the MEE of the three constraints.  First, evidently
\begin{enumerate}
\item $\int_{-\infty}^K f_{MEE}(x)\,\mathrm{d}x =    \epsilon$;
\item $\int_{-\infty}^K xf_{MEE}(x)\,\mathrm{d}x = \epsilon \nu_-$;
\item $\int_{K}^{\infty} xf_{MEE}(x)\,\mathrm{d}x = (1-\epsilon) \nu_+$.
\end{enumerate}
Hence the constraints are satisfied.  Second, $f_{MEE}$ has
an exponential form in the constraint functions, i.e., is of the form
\[  f_{MEE}(x) = C^{-1} \exp{-\left[ \lambda_1x + \lambda_2 I_{(x\leq K)} +
\lambda_3 xI_{(x \leq K)} \right]} \]
where $C=C(\lambda_1,\lambda_2,\lambda_3)$ is the normalizing constant.
(This form comes from differentiating an appropriate functional
$J(f)$ based on entropy, and forcing the integral to be unity and imposing
the constraints with Lagrange multipliers.)

The shape of $f_-$ depends on the relationship between
$K$ and the expected shortfall $\nu_-$.  The closer $\nu_-$ is to $K$, the
more rapidly the tail falls off. As $\nu_- \rightarrow K$, $f_{-}$ converges to
a unit spike at $x=K$.

\subsection{Case B: Constraining the Absolute Mean}

If instead we constrain the absolute mean, namely\\
\[ E|X|=\int |x|f(x)\,\mathrm{d}x=\mu, \]
then the MEE is somewhat less apparent but can still be found.
Define $f_-(x)$ as above, and let

\begin{equation*}
f_{+}(x) =
\begin{cases}
\frac{\lambda_1}{2-\exp (\lambda_1K)} \exp (-\lambda_1|x|)
& \text{if } x \geq K,\\
0 & \text{if } x < K.
\end{cases}
\end{equation*}
Then $\lambda_1$ can be chosen such that
\[  \epsilon \nu_- + (1-\epsilon) \int_{K}^{\infty} |x| f_{+}(x)\,\mathrm{d}x = \mu.\]

\begin{figure}
\includegraphics[width=\linewidth]{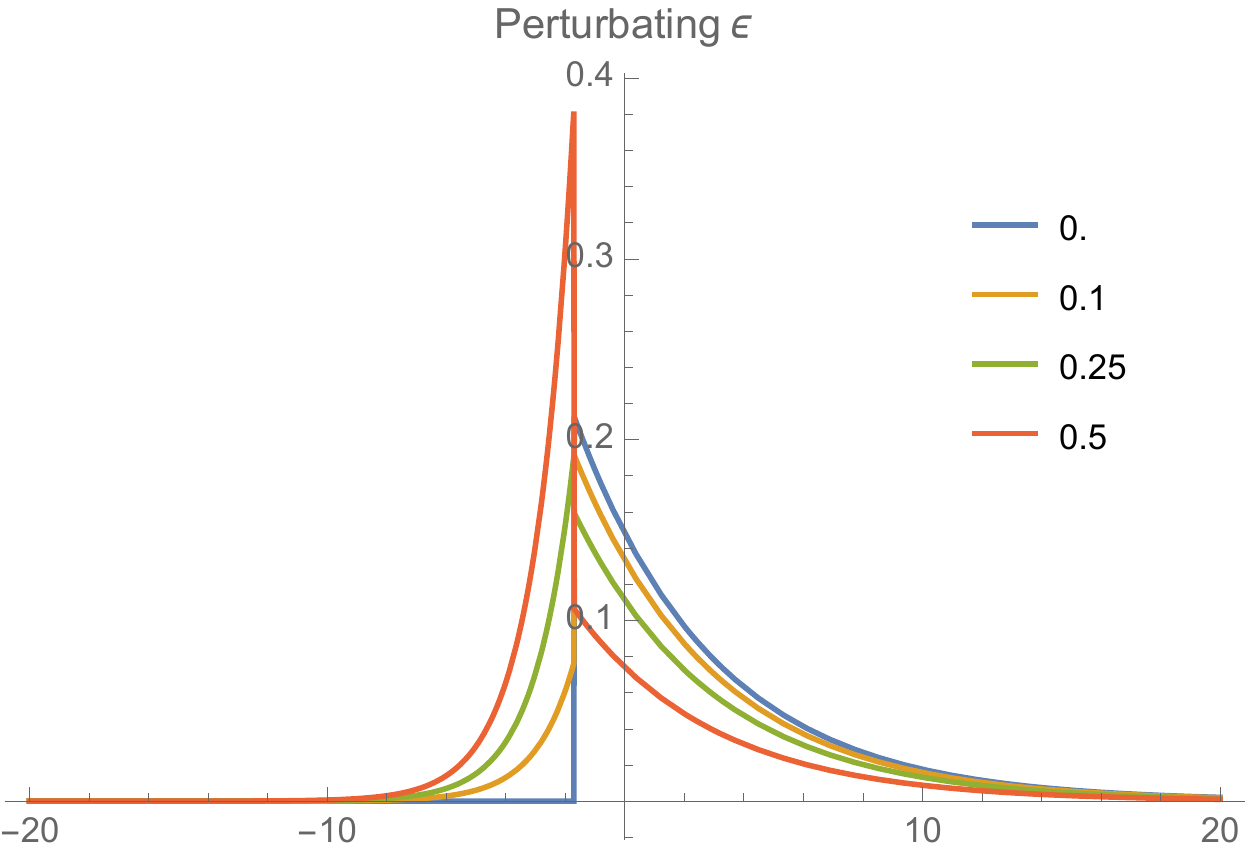}
\caption{Case A: Effect of different values of $\epsilon$ on the shape of the distribution.}
\end{figure}

\begin{figure}
\includegraphics[width=\linewidth]{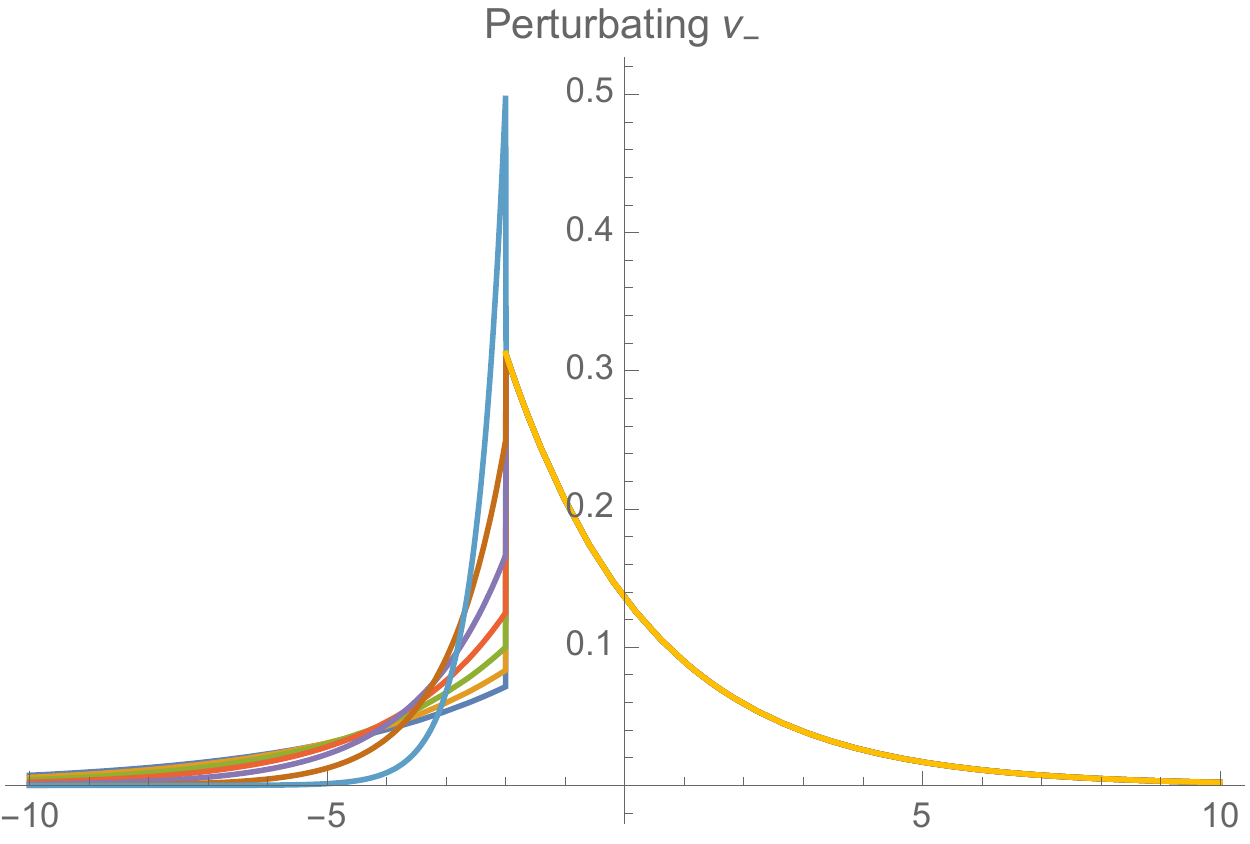}
\caption{Case A: Effect of different values of $\nu_-$ on the shape of thedistribution.}
\end{figure}

\subsection{Case C: Power Laws for the Right Tail}

If we believe that actual returns have ``fat tails,'' in particular that 
the right tail decays as a power law rather than exponentially (as with
a normal or exponential density), than we can
add this constraint to the VaR constraints instead of working with the
mean or absolute mean.  In view of the exponential form of the MEE,
the density $f_+(x)$ will have a power law, namely
\[  f_+(x)=\frac{1}{C(\alpha)} (1+|x|)^{-(1+\alpha)}, x \geq K, \]
for $\alpha >0$ if the constraint is of the form
\[ E \left(\log(1+|X|)|X>K \right)=A.\]
Moreover, again from the MEE theory, we know that the parameter is obtained by
minimizing the logarithm of the normalizing function.  In this case,
it is easy to show that
\[  C(\alpha) \doteq \int_K^{\infty}  (1+|x|)^{-(1+\alpha)}\,\mathrm{d}x = \frac{1}{\alpha}(2-(1-K)^{-\alpha}).\]
It follows that $A$ and $\alpha$ satisfy the equation
\[  A = \frac{1}{\alpha} - \frac{\log(1-K)}{2(1-K)^{\alpha}-1}.\]
We can think of this equation as determining the decay rate $\alpha$
for a given $A$ or, alternatively, as determining the constraint value $A$ 
necessary to obtain a particular power law $\alpha$.

\begin{figure}[h]
\includegraphics[width=\linewidth]{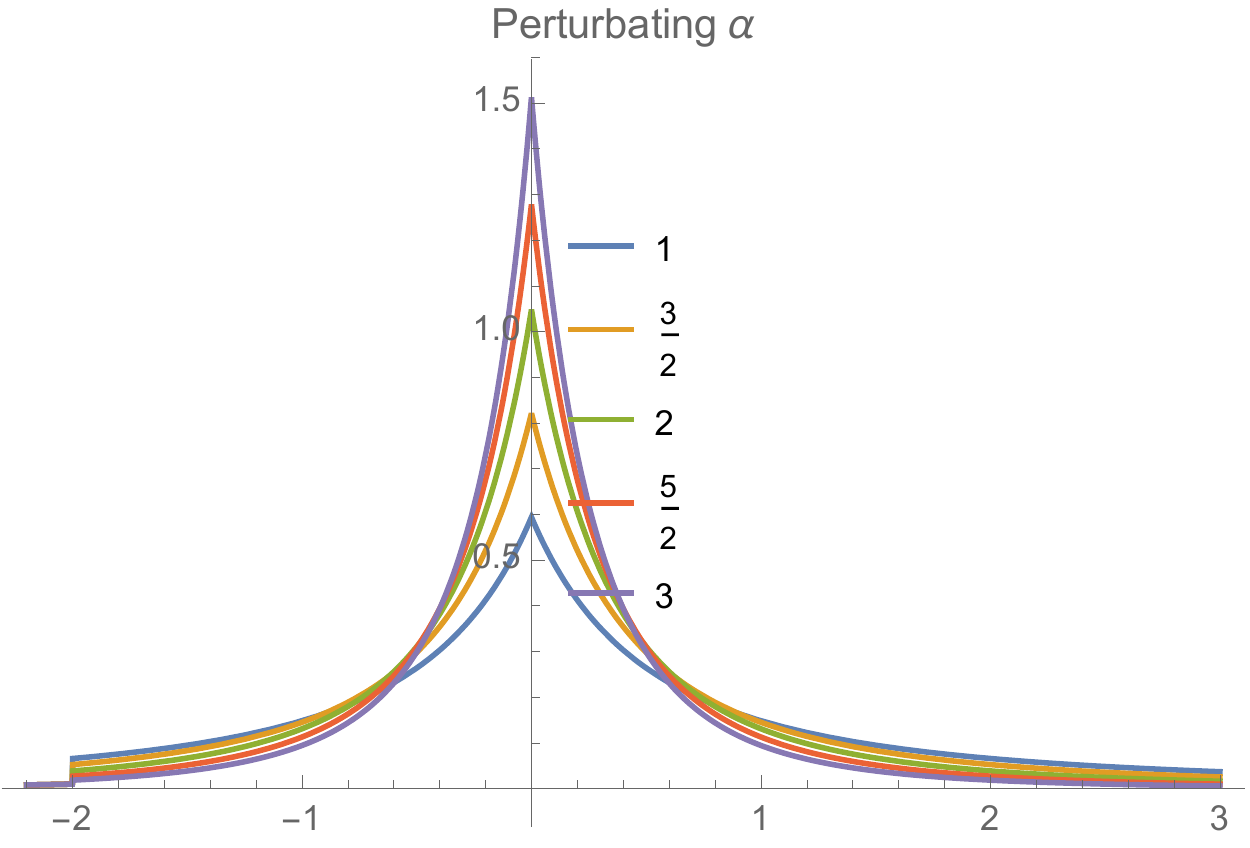}
\caption{Case C: Effect of different values of on the shape of the fat-tailed maximum entropy distribution.}
\end{figure}

\begin{figure}[h]
\includegraphics[width=\linewidth]{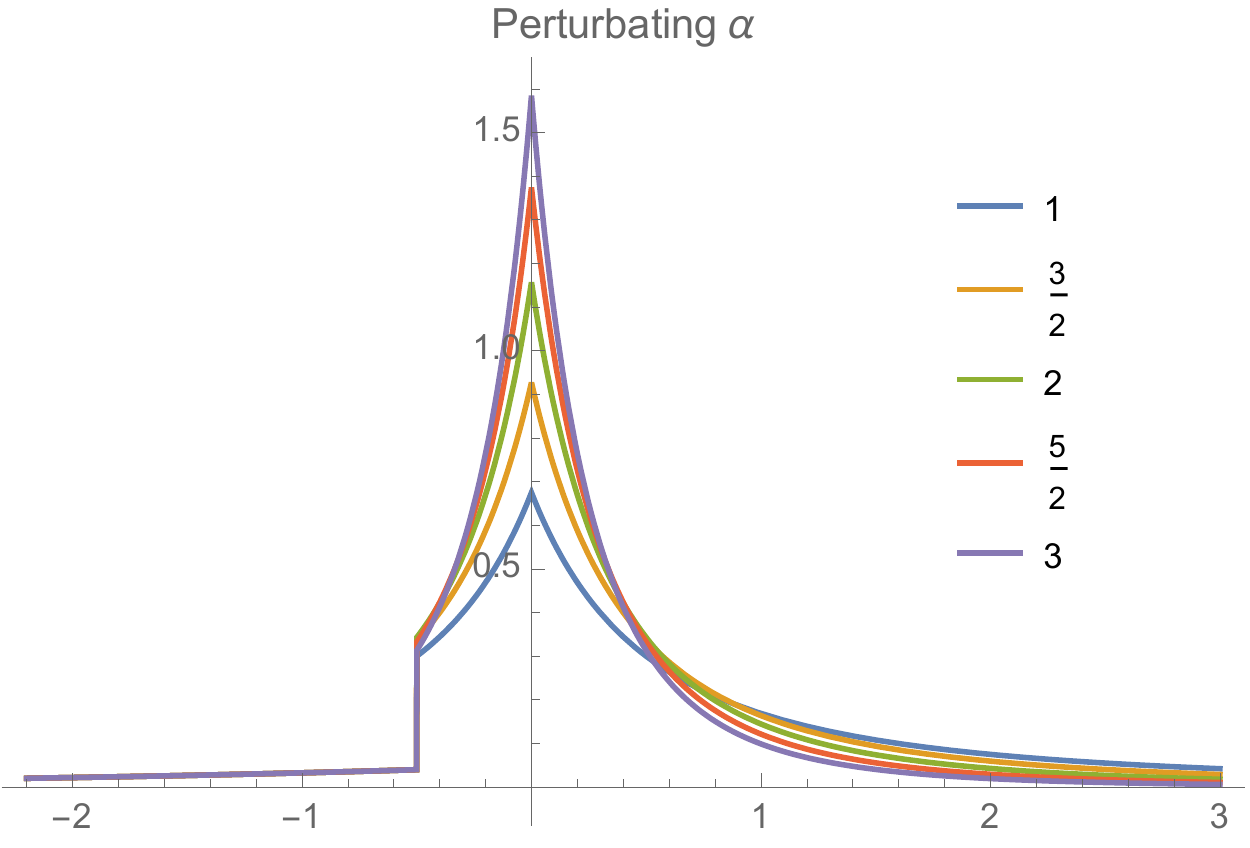}
\caption{Case C: Effect of different values of on the shape of the fat-tailed maximum entropy distribution.}
\end{figure}

The final MEE extension of the VaR constraints together with the
constraint on the log of the return is then:
\begin{multline*}
f_{MEE}(x) = \epsilon I_{(x \leq K)}\frac{1}{(K-\nu_-)} \exp 
\left[ -\frac{K-x}{K-\nu_-} \right] \\
+ (1-\epsilon)I_{(x>K)}
\frac{(1+|x|)^{-(1+\alpha)}}{C(\alpha)}.
\end{multline*}

\subsection{Extension to a Multi-Period Setting: A Comment}

Consider the behavior in multi-periods. Using a naive approach, we sum
up the performance as if there was no response to previous returns. We
can see how Case A approaches the regular Gaussian, but not Case C.
\begin{figure}
\includegraphics[width=\linewidth]{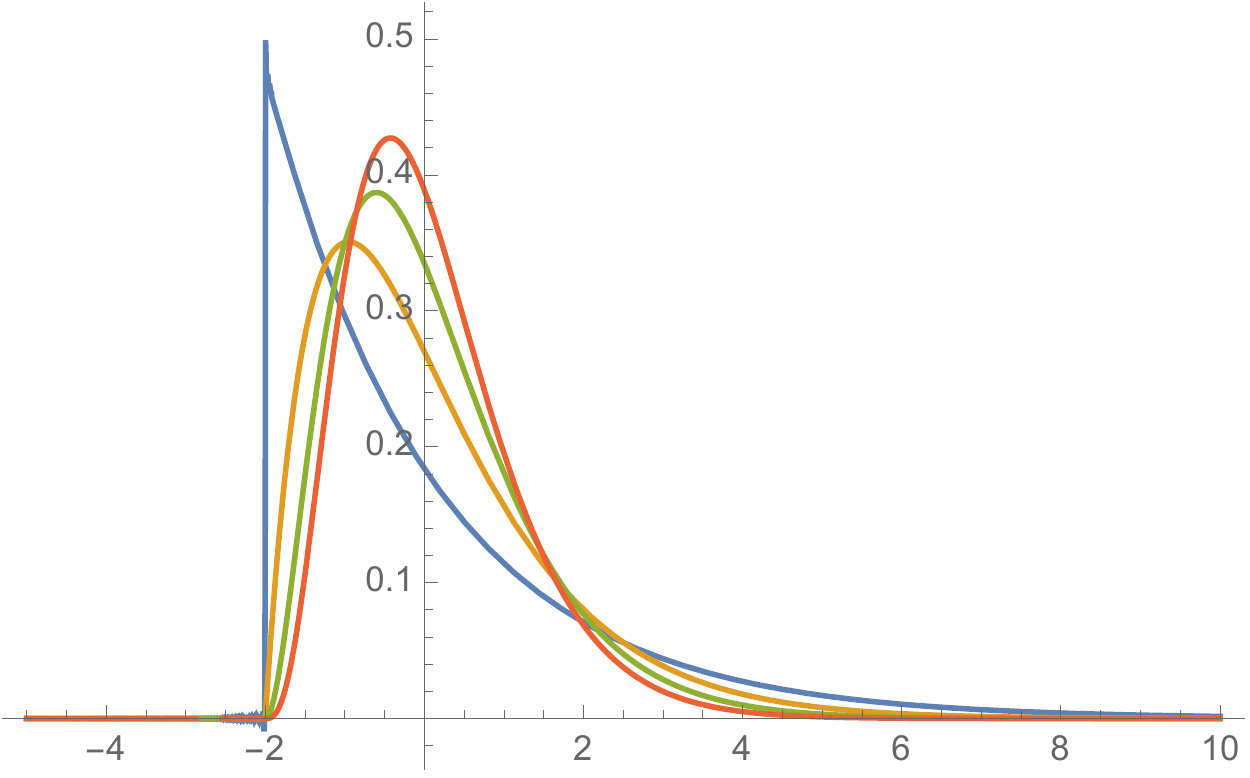}
\caption{
Average return for multiperiod naive strategy for Case A, that is,
assuming independence of "sizing", as position size does not depend on
past performance. They aggregate nicely to a standard Gaussian, and
(as shown in Equation \ref{limitaverage}), shrink to a Dirac at the
mean value.}
\end{figure}

For case A the characteristic function can be written:
$$\Psi^A(t)=\frac{e^{i K t} (t (K-\nu_- \epsilon +\nu_+ (\epsilon
-1))-i)}{(K t-\nu_- t-i) (-1-i t (K-\nu_+))}$$
   
So we can derive from convolutions that the function $\Psi^A(t)^n$ converges to that of an n-summed Gaussian. Further the 
characteristic function of the limit of the average of strategies, namely 
\begin{equation}
\lim_{n\to \infty } \,\Psi^A(t/n)^n= e^{i t (\nu_++\epsilon  (\nu_- -\nu_+))},
\label{limitaverage}
\end{equation}
is the characteristic function of the Dirac delta, visibly the effect of the law of large numbers delivering the same result as the Gaussian with mean $nu_++\epsilon  (\nu_- -\nu_+)$ .

As to the power law in Case C, convergence to Gaussian only takes
place for $\alpha \geq 2$, and rather slowly.


\section{Comments and Conclusion}    

We note that the stop loss plays a larger role in determining the
stochastic properties than the portfolio composition. Simply, the
stop is not triggered by individual components, but by variations in
the total portfolio. This frees the analysis from focusing on
individual portfolio components when the tail --via derivatives or
organic construction-- is all we know and can control.
 
To conclude, most papers dealing with entropy in the mathematical
finance literature have used minimization of entropy as an
optimization criterion. For instance, Fritelli (2000) 
\cite{frittelli2000minimal} exhibits the unicity of a "minimal entropy
martingale measure" under some conditions and shows that minimization
of entropy is equivalent to maximizing the expected exponential
utility of terminal wealth. We have, instead, and outside any utility
criterion, proposed entropy maximization as the recognition of the
uncertainty of asset distributions. Under VaR and Expected Shortfall
constraints, we obtain in full generality a "barbell portfolio" as the
optimal solution, extending to a very general setting the approach of
the two-fund separation theorem.

\bibliographystyle{IEEEtran}
\bibliography{/Users/nntaleb/Dropbox/Central-bibliography}

\section*{Appendix}

\noindent
{\bf Proof of Proposition 1:}  Since $X \sim N(\mu,\sigma^2)$,
the tail probability constraint is
\[  \epsilon = \mathbb{P}(X<K)=\mathbb{P}(Z<\frac{K-\mu}{\sigma})=\Phi(\frac{K-\mu}{\sigma}).\]
By definition, $\Phi(\eta(\epsilon))=\epsilon.$  Hence,
\begin{equation}
  K=\mu+\eta(\epsilon)\sigma 
  \label{1stapp}
  \end{equation}
For the shortfall constraint,
\begin{eqnarray*}
  \mathbb{E}(X;X<k) & = & \int_{-\infty}^K \frac{x}{\sqrt{2\pi}\sigma}
\exp-{\frac{(x-\mu)^2}{2\sigma^2}}\,\mathrm{d}x \\
         & = & \mu \epsilon 
+ \sigma \int_{-\infty}^{(K-\mu)/\sigma)} x\phi(x)\,\mathrm{d}x \\
         & = & \mu \epsilon  
-\frac{\sigma}{\sqrt{2\pi}} \exp-{\frac{(K-\mu)^2}{2\sigma^2}}
\end{eqnarray*}
Since, $\mathbb{E}(X;X<K)=\epsilon\nu_-$, and from the definition of
$B(\epsilon)$, we obtain
\begin{equation}
\nu_- = \mu - \eta(\epsilon) B(\epsilon)\sigma \label{2ndapp}
\end{equation}

Solving (\ref{1stapp}) and (\ref{2ndapp}) for $\mu$ and $\sigma^2$ gives the expressions
in Proposition 1.  

Finally, by symmetry to the ``upper tail inequality''
of the standard normal, we have, for $x <0$,
$\Phi(x) \leq \frac{\phi(x)}{-x}$ for $x>0$. Choosing
$x=\eta(\epsilon)=\Phi^{-1}(\epsilon)$ yields
$\epsilon = P(X < \eta(\epsilon)) \leq -\epsilon B(\epsilon)$ or
$1+B(\epsilon)\leq 0$.  Since the upper tail inequality is
asymptotically exact as $x \rightarrow \infty$ we have
$B(0)=-1$, which concludes the proof.



\end{document}